\documentclass[prb,twocolumn]{revtex4}
\usepackage{graphicx}
\begin{document}


\title{Resistivity and Thermopower of Ni$_{2.19}$Mn$_{0.81}$Ga\\}

\author{K. R. Priolkar, P. A. Bhobe, Shannol Dias Sapeco and Rajkumar Paudel}
 \affiliation{Department of Physics, Goa University, Goa, 403 206 India.}

\date{\today}

\begin{abstract}
In this paper, we report results of the first studies on the thermoelectric power (TEP) of the magnetic heusler alloy Ni$_{2.19}$Mn$_{0.81}$Ga. We explain the observed temperature dependence of the TEP in terms of the crystal field (CF) splitting and compare the observed behavior to that of the stoichiometric system Ni$_2$MnGa. The resistivity as a function of temperature of the two systems serves to define the structural transition temperature, T$_M$, which is the transition from the high temperature austenitic phase to the low temperature martensitic phase. Occurrence of magnetic (Curie-Weiss) and the martensitic transition at almost the same temperature in Ni$_{2.19}$Mn$_{0.81}$Ga has been explained from TEP to be due to changes in the density of states (DOS) at the Fermi level.  
\end{abstract}
\maketitle

\section{\label{sec:level1}Introduction}
Ni$_2$MnGa is one of the shape memory effect compound which is currently exciting and has gained considerable interest since it is ferromagnetic. The origin of shape memory effect in Ni$_2$MnGa is in the martensitic transition which takes place on cooling through 220K from the cubic L2$_1$ Heusler structure to a tetragonal phase. If the material is plastically deformed in the low temperature martensitic phase and the external load removed it regains its original shape when heated above the transition temperature. Based on early neutron diffraction data the transformation has been described as a simple contraction along the \{100\} direction of the cubic cell without any change in atomic positions \cite{Web}. This phase transition is remarkable in that, in spite of the large deformation, it is reversible and a single crystal can be cycled through it many times without breaking. In recovering their shape the alloys can produce a displacement or a force, or a combination of the two, as a function of temperature. Because of these novel and remarkable properties shape memory alloys find themselves in large number of applications in the fields of engineering and medicine \cite{mvg}. Since Ni$_2$MnGa orders ferromagnetically below 375K the possibility of producing giant field induced strains, which are an order of magnitude larger than those observed in rare-earth transition metal alloys, has stimulated large number of investigations \cite{proc}.

Recently it has been found that in Ni-Mn-Ga systems huge strains can be induced by application of magnetic field \cite{ulla,rc,tick,soz}. These compounds undergo a martensitic transformation between a low temperature tetragonal phase which is magnetically hard and a high temperature cubic phase (magnetically soft) \cite{Web,alb}. This difference in the anisotropy strongly modifies the field dependence of the magnetization in the two phases, with the saturation magnetization value being slightly lower in the cubic austenite \cite{teg,hu1}. Some recent works have evidenced the occurrence of significant isothermal variations of the magnetic entropy in NiMnGa compounds (up to $|\Delta S_m|$ = 18 J/kg-K for H = 5 T) in correspondence to the martensitic transformation. In these cases, the martensitic transition temperatures T$_M$ are lower than the Curie temperature T$_C$ and, as a consequence, the martensitic transformation takes place between two ferromagnetic phases \cite{teg,hu1,hu2}. For a composition for which ${\rm T}_C \sim {\rm T}_M$  occurrence of  large magneto-caloric effect has been demonstrated recently \cite{par}.

Studies on Ni$_{2+x}$Mn$_{1-x}$Ga alloys have emerged with a phase diagram which indicate that partial substitution of Mn for Ni results in the increase in the structural phase transition temperature T$_M$ (martensitic transition) and the decrease in the Curie temperature T$_C$ up to their coincidence at x $\approx$ 0.19 \cite{vasil,boz}. Theoretical analysis demonstrating the importance of the conduction electron density in stabilizing the Heusler structure was noted a long ago and the suggestion that the structure is stabilized because the Fermi surface touches the Brillouin zone boundary was made by \cite{Smit}.
The aim of the present work is to investigate the transport properties of the polycrystalline Ni$_{2.19}$Mn$_{0.81}$Ga alloy and compare it with the stoichiometric Ni$_2$MnGa alloy in order to understand the effect of excess Ni on the DOS at Fermi level. For this purpose we have studied the temperature dependent resistivity and thermoelectric power (TEP) which is sensitive to the changes in the DOS at the Fermi level of the two alloys.  

\section{\label{sec:level1}Experimental}
\begin{figure}[h]
\includegraphics{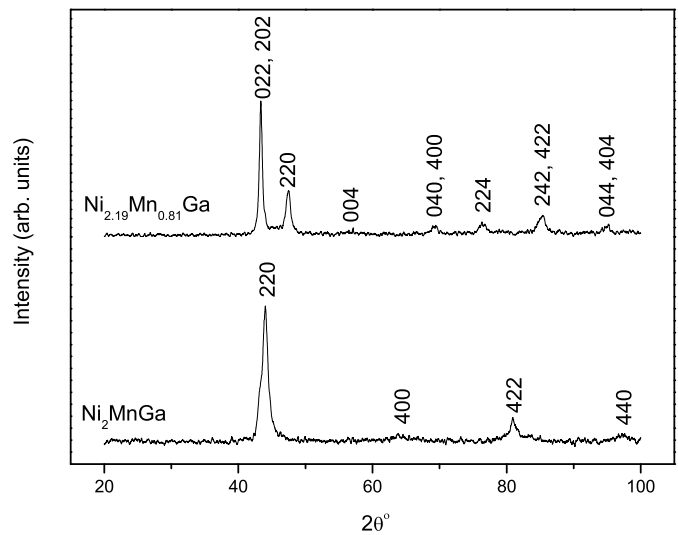}
\caption{\label{xrd} X-ray diffraction patterns of Ni$_2$MnGa and Ni$_{2.19}$Mn$_{0.81}$Ga}
\end{figure}

Polycrystalline Ni$_{2+x}$Mn$_{1-x}$Ga (x = 0, 0.19) ingots were prepared by the conventional arc-melting method in argon atmosphere. The starting materials with 4N purity were taken in the stoichiometric ratio and were remelted 4-5 times to attain good compositional homogeneity. Since the weight loss during melting was approximately $\le$ 0.5 \%, the composition of the ingots was assumed to be nominal. X-ray diffraction powder pattern recorded in the range $ 20^\circ \le 2\theta \le 100^\circ$ confirmed that the samples were homogeneous and of single phase with no detectable impurity and the patterns are presented in Fig.\ref{xrd}. The Ni$_2$MnGa has a cubic L2$_1$ structure at room temperature with lattice parameter $a$ = 5.824\AA.~ As the martensitic transition temperature for Ni$_{2.19}$Mn$_{0.18}$Ga is $\sim$ 320K, the XRD pattern represents a structure with lower symmetry. This pattern can be indexed to a body centered tetragonal structure ($I_4/mmm$)\cite {wir} or to a face centered orthorhombic structure ($Fmmm$) with a = b \cite{Web}. It may be noted here that the tetragonal and orthorhombic structural models are related to each other by a simple transformation matrix \cite{wedel}. In the orthorhombic structural model only the lattice parameters change from that of cubic high temperature phase but the relative atom coordinates remain unchanged.  The lattice parameters obtained from the orthorhombic model were $a$ = 5.416\AA~ and $c$ = 6.523\AA.~ 

Electrical resistivity was measured using the standard four-probe technique. The samples were first cooled to 80K and the resistance was measured upon warming upto 350K followed by subsequent cooling back to 80K. 
The thermopower measurements were carried out using the differential method where the voltage difference $\Delta$V due to temperature difference $\Delta$T across the sample was measured in the temperature range 100K to 400K in the warming/cooling cycles similar to that of resistivity measurements. The sample was kept between two highly polished copper plates, electrically insulated from the rest of the sample holder. Two heater coils, one on the bottom and the other on the top copper plate, served to raise the overall temperature of the sample and to maintain a temperature gradient across the length of the sample respectively. The overall temperature of the sample was measured by a Platinum Resistance thermometer (PT-100) while the gradient was monitored by a copper-constantan thermocouple operating in the differential mode. To measure the thermopower $S$ at a particular temperature say $T$, the temperature difference across the sample is first adjusted to nearly 0 K ($\sim 1\mu V$) by passing current through the two heater coils. The top copper plate of the sample holder is then heated resulting in a thermo emf  ${V_{s}}$ across the sample. The voltages $V_s$ and that developed across the thermocouple  ${V_{th}}$ are measured for different temperature gradients between the two plates. A graph of ${V_{s}}$ versus ${V_{th}}$ is plotted and its slope ($\Delta V_{s}/ \Delta V_{th}$) is measured. Knowing the slope and the thermopower, ${S_{th}}$ of the thermocouple at $T$, thermopower $S$ is obtained.
 
\section{\label{sec:level1}Results}
\subsection{\label{sec:level2}Resistivity}
\begin{figure}
\includegraphics{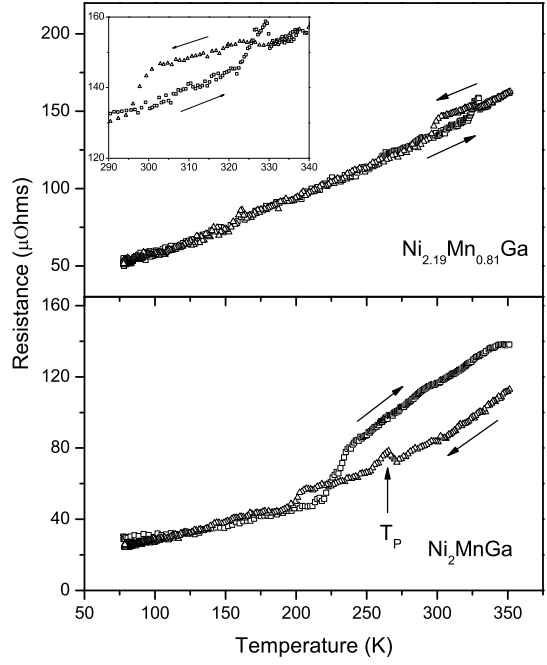}
\caption{\label{res} Plots of resistance versus temperature for Ni$_2$MnGa and Ni$_{2.19}$Mn$_{0.81}$Ga}
\end{figure}

 The temperature dependencies of resistivity measured upon warming and cooling in Ni$_2$MnGa and Ni$_{2.19}$Mn$_{0.81}$Ga are presented in Fig.\ref{res}. A large hysteresis is observed upon thermal cycling in both the compositions as evident from Fig.\ref{res} and the inset therein. This could be due to the variation in the percentage conversion from a five-layered modulation (5M) to seven-layered modulation (7M) termed as inter-martensitic transition occuring at low temperatures \cite{khov2}. In such a transition, the sample undergoes a transformation from 5M state to 7M state upon cooling. This transformation depends on the warming/cooling rate and in a given experimental condition a complete conversion may not be achived. Upon subsequent heating, the reverse transformation i.e. 7M $\rightarrow$ 5M is absent and this leads to different behavior of transport properties upon warming and cooling.  

On warming, Ni$_2$MnGa exhibits a jump-like feature at around 210K which is associated with a transition from the martensitic to the austinitic phase for this alloy. Cooling from high temperature austinitic phase results in a well-defined peak at aroung 265K which marks the pre-martensitic transition (T$_P$) in agreement with Khovailo {\em et al} \cite{khov}. As reported in the literature the ferromagnetic transition for this alloy takes place at T$_C \sim$ 380K which is beyond the studied temperature range and hence we do not observe any such signature in the resistance measurement for this alloy. 
In Ni$_{2.19}$Mn$_{0.81}$Ga, the structural transition T$_M$ (320K) and the ferromagnetic transition T$_C$ (322K) occur at almost the same temperatures. Moreover these transitions being very broad, the pre-martensitic transition as observed in Ni$_2$MnGa is not revealed in this alloy. Also, theoretical as well as experimental studies on Ni-Mn-Ga indicates that the pre-martensitic transformation is observed only in the alloys with T$_M < $ 260K \cite{cas,pla,gon,stu}. Thus absence of the anomaly assigned as T$_P$ in the Ni$_{2.19}$Mn$_{0.81}$Ga alloy is expected due to its high martensitic transition temperature.

\subsection{\label{sec:level2}Thermopower}
\begin{figure}
\includegraphics{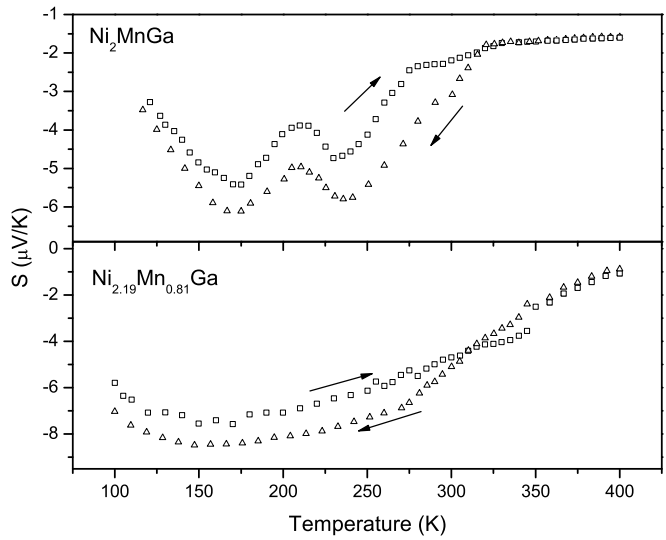}
\caption{\label{tep1} Thermoelectric power as a function of temperature for Ni$_2$MnGa and Ni$_{2.19}$Mn$_{0.81}$Ga}
\end{figure}

Thermopower is very sensitive to the energy dependence of the carrier mobility near the Fermi level which in turn depends on the crystal structure concerned. Hence, TEP of Ni$_{2.19}$Mn$_{0.81}$Ga alloy would be expected to exhibit interesting behavior in the vicinity of the austenitic to martensitic phase transition given that the two different crystal structures are involved along with the ferromagnetic transition occurring at the same temperature. Fig.\ref{tep1} shows the temperature dependencies of TEP for the two alloys in the temperature range 100K to 400K. The striking feature is the broad hump obtained in the vicinity of T$_M$ for Ni$_2$MnGa due to the austinitic to martensitic transition. Such a feature is absent for Ni$_{2.19}$Mn$_{0.81}$Ga. As the fact that the T$_C$ for Ni$_2$MnGa and Ni$_{2.19}$Mn$_{0.81}$Ga are $\sim$ 380K and $\sim$ 322K respectively, there is a contribution from the magnetic scattering to the TEP in the austinitic phase which is clearly evident from the steep fall of $S$ with the decrease in temperature in this region. As the temperature is further decreased $|S|$ shows a strong negative dip in TEP and finally resumes the normal metallic behavior of S$\rightarrow$ 0 as T$\rightarrow$ 0. Such a behavior is typical of the Heusler alloys representing the fact that they are the good approximation towards the local-moment ferromagnetic systems \cite{camp}. The negative dip occurs at T$\sim$ 0.4T$_C$ and a weak temperature dependence of TEP around Curie temperature is also seen as observed for other Heusler alloys \cite{camp}.

The warming data for TEP is lower in magnitude in comparision with the subsequent cooling results. This as mentioned above can be explained to be due to the 5M $\rightarrow$ 7M intermartensitic transition occuring due to thermal cycling of the alloys during measurements.      

To explain the observed anomalies in $S$, we consider two scattering contributions to TEP, the magnetic scattering of the thermal current as both the alloys are magnetically ordered at lower temperatures and the structural (martensitic) transition scattering. Thus the total TEP can be written as 
\[ S = S_m + S_s\]
where \[S_m = \alpha \times T^{\frac{3}{2}}\]
is the magnetic contribution and \[S_s = -\frac{1}{e \sigma T} \int (\epsilon - \mu)\frac{\partial f_0}{\partial \epsilon}\sigma(\epsilon)d\epsilon\]
where $f_0$ is the Fermi-Dirac distribution function,~ $\mu$ is the chemical potential and \[ \sigma = \int \sigma(\epsilon)\frac{\partial f_0}{\partial \epsilon}d\epsilon\]
which is the simple semi-classical result for thermal diffusion in metallic systems.

Owing to the fact that there is not much change in the atom positions in transition from cubic to tetragonal structure, a safe assumption that any change in the TEP is a direct manifestation of the changes in the density of states (DOS) can be made.  
 
\begin{figure}
\includegraphics{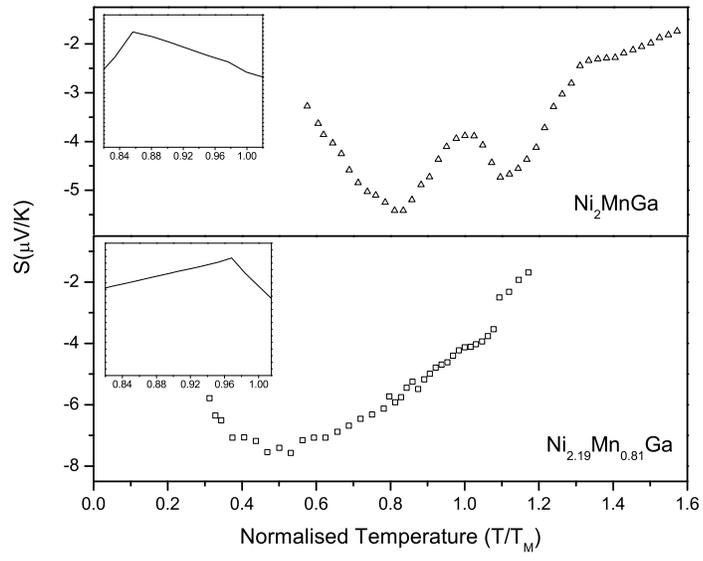}
\caption{\label{tep2} Thermoelectric power as a function of normalized temperature for Ni$_2$MnGa and Ni$_{2.19}$Mn$_{0.81}$Ga}
\end{figure}

Fig.\ref{tep2} represents the TEP data with the temperature axis normalized with respect to the matensitic temperatures of the respective alloys. The TEP data in the 0.80 $\le$ T/T$_M \le$ 1.02 range shows an inflection point at normalized temperature of $\sim$ 0.86 and $\sim$ 0.94 (see inset) for Ni$_2$MnGa and Ni$_{2.19}$Mn$_{0.81}$Ga respectively. If a model for DOS near Fermi level is assumed consisting of a peak near Fermi level, the the TEP of both the alloys can be accounted for by this peak shifiting closer to the Fermi level in Ni$_{2.19}$Mn$_{0.81}$Ga. The shift of the inflection point from 0.86 to 0.94 can then be associated to the shift of peak in DOS towards the Fermi level as the Ni content is increased in going from Ni$_2$MnGa to Ni$_{2.19}$Mn$_{0.81}$Ga. 

\section{\label{sec:level1}Discussion}

Increasing Ni doping for Mn results in doping electrons in the DOS at Fermi level. This considerably alters the band structure as is evident from the change in magnitude of thermopower in proceeding from the martensitic to austenitic phase. The decrease in magnetic ordering temperature coupled with increase in martensitic transition temperature can also be understood from here. Nickel with nearly full 3d band when replaces Mn with nearly half filled band, results in reduction of magnetic moment and therefore the magnetic transition temperature. Similarly, change in the position of the peak in DOS at Fermi level, which is associated with Ni 3d band results in phase instability and therefore a phase transition from cubic austentic phase to tetragonal martensitic phase. Such peaks in electronic DOS are known to lead to a structural phase transition \cite{mark}.
These changes are probably a result of a redistribution of 3d electrons amongst the 3d orbitals whose degeneracy is further broken by the lowered symmetry from cubic to tetragonal. Substitution of Ni for Mn results in transfer of electrons from the nearly full 3d band of nickel to more than half filled 3d band of manganese. It is the splitting of energy sub-bands which are degenerate in the cubic phase which enables the electrons to redistribute themselves so as to lower the free energy. This is the well known band Jahn-Teller mechanism. In the band model there is an increase in the width of the energy bands because, when the crystal deforms there is a change in the degree of overlap of the associated orbitals. Unlike in the case of stoichiometric Ni$_2$MnGa, where the c/a ratio of the tetragonal phase is $<$ 1, for Ni$_{2.19}$Mn$_{0.81}$Ga, the c/a ratio is $>$ 1. This will lead to a redistribution of electrons in the crystal field split 3d band of this alloy. A redistribution of magnetization is found for stoichiometric Ni$_2$MnGa in the neutron scattering experiment as a function of temperature when it undergoes a transition from high temperature austinitic to low temperature martensitic phase \cite{brow}.

The band structure of Ni$_2$MnGa has been calculated by \cite{fuj}. The composition of bands that are active at the Fermi surface could be identified. With this identification the Fermi level lies just above a peak in the DOS of the minority spin Ni $e_g$ band and at a position in the Mn band there is an almost equal DOS of majority and minority spin $t_{2g}$ states. For a martensitic transition to occur important feature required is that the peak in the DOS should have some asymmetry, whereby it has more weighting towards lower energies and that the Fermi level is situated very close to the peak. Such a DOS can explain the observed thermopower very well, especially in the martensitic phase and has been used to explain thermopower data of shape memory NiTi alloys \cite{lee}. In the present study, thermopower can be explained by assuming a similar model of DOS. A shift in the position of the peak in DOS is observed towards Fermi level with increasing Ni concentration. The assumed model, on integration, yeilds the same variation as TEP observed experimentally in the present study.

\section{\label{sec:level1}Conclusion}
In conclusion, we have studied the resistivity and investigated the variation in thermopower for the magnetic heusler alloys, Ni$_2$MnGa and Ni$_{2.19}$Mn$_{0.81}$Ga. The experimental results indicate that the TEP for Ni$_{2.19}$Mn$_{0.81}$Ga though different from that for Ni$_2$MnGa, the general trend in the variation of TEP is that of a typical Heusler alloy with local-moment ferromagnetism. With the assumed model, the peak in DOS just below Fermi level is seen to shift towards higher energy in the region of the martensitic transition with increasing Ni content. All the anomalies observed in TEP have been explained to be due to crystal field splitting and the associated changes in the density of states near the Fermi level.

\acknowledgments
K. R. Priolkar would like to  acknowledge financial support from Council for Scientific and Industrial Research under the project 03(0894)/99/EMR-II. P. A. Bhobe thanks Inter University Consortium for DAE Facilities, Mumbai for the fellowship. Authors thank Prof. P. R. Sarode for his constant encouragement and guidance.

\newpage

\end{document}